# Performance Analysis of Noise Subspace-based Narrowband Direction-of-Arrival (DOA) Estimation Algorithms on CPU and GPU

*Hamza Eray, Alptekin Temizel*

Graduate School of Informatics
Middle East Technical University
`{eray.hamza, atemizel}@metu.edu.tr`

*Abstract*— High-performance computing of array signal processing problems is a critical task as real-time system performance is required for many applications. Noise subspace-based Direction-of-Arrival (DOA) estimation algorithms are popular in the literature since they provide higher angular resolution and higher robustness. In this study, we investigate various optimization strategies for high-performance DOA estimation on GPU and comparatively analyze alternative implementations (MATLAB, C/C++ and CUDA). Experiments show that up to 3.1x speedup can be achieved on GPU compared to the baseline multi-threaded CPU implementation. The source code is publicly available at the following link: *https://github.com/erayhamza/NssDOACuda*

## 1. Introduction

Direction-of-arrival (DOA) estimation methods, which fall under the field of array signal processing, have been studied since the early 1900s with the usage of directional antenna element (loops, dipoles, etc.) characteristics [1] and various approaches have been proposed [2]. These estimation methods are widely used in several civil/military applications such as radar, sonar, passive source location, and wireless communication [3]. Search-and-rescue (beacon) [4] and military signal intelligence by passive electromagnetic (EM) direction finding (DF) are the most critical applications of DOA estimation.

This study is based upon the assumption that there are more than one virtual uncorrelated RF sources synthetically generating narrowband signals impinging on an antenna array with a specific number of elements and geometry. The main purpose is to estimate DOA of these sources with sufficient accuracy within a certain duration. For that purpose, amongst several methods in the DOA-estimation literature, noise subspace-based methods are chosen as they provide higher angular-resolution and more robust estimation [1].

The subspace-based approaches have started a new era in the sensor array signal processing, succeeding the classical (beamformer) methods since they provide better estimation performance [2]. The first such method was suggested by Pisarenko in 1973 [5]. PHD (Pisarenko Harmonic Decomposition) is related to the frequency estimation of complex exponential sum within the white noise. Based upon Carathéodory theorem, it was proven that the frequency information could be extracted from the eigenvector corresponding to the minimum eigenvalue of $R_{hat}$ matrix [6]. After PHD, based upon the similar idea, the MUSIC (**MU**ltiple **SI**gnal **C**lassification) algorithm [7] has become quite popular. In the MUSIC algorithm, the scope for noise subspace selection was extended and hence its usage was made more generic. EV (**E**igen **V**ector) [8] and MN (**M**inimum **N**orm) [9] algorithms are modified versions of MUSIC by some weights/norms. These four closely related algorithms (PHD, MUSIC, EV and MN) are investigated within the scope of this study.

For all these methods, we follow the following steps (*i*) Algorithm analysis, MATLAB implementation and generation of ground-truth results, (*ii*) Implementation in C/C++ and code profiling, (*iii*) Implementation and parallelization in CUDA and numerical validation, (*iv*) Performance optimization and benchmarking.

In the first step, each algorithm is studied theoretically, modeled mathematically, and implemented in the MATLAB environment. Then it is simulated with synthetic input test data. Output (power) values and estimation results are computed in double precision and they are considered as the ground truth. MATLAB code run-times are measured to use in the subsequent code performance analyses.

In the second step, each algorithm is implemented in C/C++. Corresponding output values and estimation results are compared against ground-truth values calculated in the previous step and percent error is measured. The algorithms are performance profiled to identify the time-consuming parts and the parallelization potentials of these parts are investigated. Then a multi-threaded CPU-implementation is done using OpenMP (Open Multi-Processing) [10]. The performance of these C/C++ implementations is measured under different DOA angle scan values.

In the third step, the time-consuming parts determined in the previous step are analyzed with regards to their suitability for massively parallel implementation on GPU. After this analysis, each algorithm is implemented in CUDA - C/C++ in a hybrid mode (CPU+GPU). To numerically validate CUDA implementations, the same error measurement and estimation comparison procedure in the second step is followed. The performance of these algorithms in different scan ranges and durations are compared against C/C++ and MATLAB implementations.

In the last step, CUDA implementations are optimized and performance evaluations are done on different configurations.

The paper is structured as follows: in Section 2, we describe the previous works on the parallelization of the DOA estimation algorithms. Then we provide a theoretical background regarding these aforementioned algorithms in Section 3. The implementation details and parallelization strategies are described in Section 4. Numerical validation of the algorithms and their performance evaluations are presented in Section 5.

## 2. Related Studies

Due to its popularity and general-purpose usage potential in DOA estimation, only the MUSIC algorithm was studied for parallelization among these four noise subspace-based methods. However, even for MUSIC, there are a limited number of studies that realize parallel implementations. In an early work by Zou et al. [11], the MUSIC algorithm was implemented on FPGA with high-speed parallel optimization by modifying some time-consuming parts with a little estimation performance degradation in return. Majid et al. [12] implemented the wideband MUSIC DOA algorithm on multi-core CPU, GPU, and IBM Cell BE Processor and they reported better performance on GPU compared to the other processors. They also emphasize that overall performance depends upon the balance between the number of tasks and data size. In [13], the wideband MUSIC DOA algorithm was realized with Parallel Haar Wavelet Transform (PHWT) approach and implementations were realized on CPU and GPU using data-level & instruction-level parallelism. In that study, the GPU implementation was reported to outperform the multi-core CPU implementation. In a recent work by Lu et al. [14], for DOA estimation of broadband underwater acoustic signals, MUSIC was implemented on CPU and GPU platforms with large-scale array and multiple frequency points. They reported a significant speedup ranging from 14 to 23 times for 256 array elements.

## 3. Background

### 3.1. Signal Data Model

In all analyses in this study (i.e., numerical validation and time duration experiments), synthetic input signals generated using a signal generation simulator in MATLAB were used. The simulator can be controlled by certain technical parameters such as sampling frequency, number of direct paths, and signal-to-noise ratio (SNR) in order to achieve the desired characteristics.

The data signal model is based on the settings [15] shown in Figure 1. In this mode, it is assumed that there is an array consisting of $m$ sensors with arbitrary geometry and there are $d$ point sources generating narrowband incoherent signals impinging on the array.

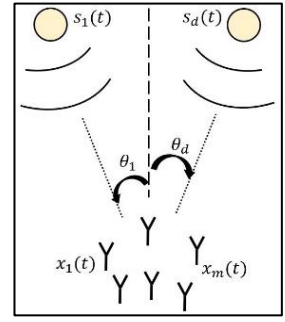

*Figure 1*: Point emitters and a passive sensor array.

In our application, these sensors are considered as omnidirectional antennas and RF signals are coming from far-field sources. Bearing in mind these assumptions, the signal model (Eq. 1) for a specific time instant $t$ is described in Table 1.

$$X(t) = A(\theta)S(t) + W(t) \quad (1)$$

*Table 1*: Signal model parameter list.

| | |
|---|---|
| Array output matrix | $X(t) = [x_1(t), \ldots, x_m(t)]^T$ |
| Incident complex signals | $S(t) = [s_1(t), \ldots, s_d(t)]^T$ |
| Zero-mean Gaussian noise | $W(t) = [n_1(t), \ldots, n_m(t)]^T$ |
| Unknown DOAs of the incident signals | $\theta = [(\theta_1, \phi_1), \ldots, (\theta_d, \phi_d)]^T$ |
| Array manifold | $A(\theta) = [a(\theta_1, \phi_1), \ldots, a(\theta_d, \phi_d)]$ |

In Table 1, $A(\theta)$ matrix is composed of $a(\theta_i, \phi_i)$, each of which is located along a matrix column and represents a steering vector corresponding to an array response due to $i^{\text{th}}$ source impinging at the array with angle pair $(\theta_i, \phi_i)$. The steering vector depends mainly upon arrival angle and antenna array geometry, and its generic expression for all geometries is given in Eq. 2.

$$a(\theta_i, \phi_i) = \begin{bmatrix} \exp\left\{j\frac{2\pi}{\lambda}\begin{pmatrix} x_1 \sin\theta_i \sin\phi_i + \\ y_1 \cos\theta_i \sin\phi_i + z_1 \cos\phi_i \end{pmatrix}\right\} \\ \vdots \\ \exp\left\{j\frac{2\pi}{\lambda}\begin{pmatrix} x_m \sin\theta_i \sin\phi_i + \\ y_m \cos\theta_i \sin\phi_i + z_m \cos\phi_i \end{pmatrix}\right\} \end{bmatrix} \quad (2)$$

In all DOA algorithms in consideration, the sample covariance matrix $\widehat{R}$ is used to store the correlation between the signals obtained from different array elements and it could be computed as in Eq. 3 where $N$ is the number of time samples.

$$\widehat{R} = \frac{1}{N}\sum_{k=1}^{N} X(t_k) X^H(t_k) = \frac{1}{N}(X_N * X_N^H) \quad (3)$$

### 3.2. Mathematical Modeling

We analyzed the similarities and common operations of these four noise subspace algorithms to facilitate code reuse. Our analysis revealed that the algorithms have many commonalities, as shown in Table 2. As can be seen from this table, while Steps 1, 2, 5, and 6 are the same, Steps 3 and 4 are different, and these steps eventually determine the resulting DOA estimation performance of different algorithms. Details of these steps for each algorithm are given in Table 3. The main difference in Step-3 and Step-4 is related to how noise subspace is spanned by which eigenvector(s) chosen. Some weighting and norm operations also play a role in the differences between the algorithms.

*Table 2*: Common code structure of four algorithms.

| Step-1 | $R_{hat} \leftarrow (X * X^h)/N$ |
|---|---|
| Step-2 | $[u, d, v] \leftarrow jsvd\ (R_{hat})$<br>$u, v$ : left/ right singular vectors<br>$d$ : diagonal singular value matrix |
| Step-3 | Noise subspace selection (NSS)<br>(different for each) |
| Step-4 | Calculation of inner product term<br>(different for each & related to Step-3) |
| Step-5 | $AngleSer = \{(\phi_i, \theta_j) \in (azSer\ x\ elevSer)\}$<br>$\mathbf{for}\ ind = 1 : len\ (AngleSer)$<br>$\quad a_{ind} \leftarrow steeringVec(\phi_i, \theta_j)$<br>$\quad a_{ind}^H \leftarrow a_{ind}.adjoint()$<br>$\quad invP_{ind} \leftarrow a_{ind}^H * C * a_{ind}$<br>$\quad PsSpc(i,j) \leftarrow P_{ind} \leftarrow (invP_{ind})^{-1}$<br>$\mathbf{end}$ |
| Step-6 | $(peakVals, inds) \leftarrow findPeaks(PsSpc)$<br>$(azDOA, elevDOA) \leftarrow PeakSelection(peakVals, inds, D)$ |

*Table 3*: Code differences for Step-3 & Step-4.

| | | |
|---|---|---|
| **Step-3** | PHD | $minIdx \leftarrow \min(d.diag)$<br>$e_{min} \leftarrow u.col(minIdx)$ |
| | MUSIC | $E_n \leftarrow u.col(D+1:end)$ |
| | EV | $E_n \leftarrow u.col(D+1:end)$<br>$w_{dn} \leftarrow d.diag(D+1:end)$<br>$E_{n,w} \leftarrow E_n.cols * (1/w_{dn})$ |
| | MN | $E_n \leftarrow u.col(D+1:end)$<br>$P_n \leftarrow E_n * E_n^H$<br>$unitV \leftarrow [1,0,...,0]_m$<br>$\lambda \leftarrow (unitV^H * P_n * unitV)^{-1}$<br>$valph \leftarrow P_n * \lambda * unitV$ |
| **Step-4** | PHD | $C \leftarrow e_{min} * e_{min}^H$ |
| | MUSIC | $C \leftarrow E_n * E_n^H$ |
| | EV | $C \leftarrow E_{n,w} * E_n^H$ |
| | MN | $C \leftarrow valph * valph^H$ |

## 4. Implementations

In order to experimentally evaluate numerical accuracy and their computational performance, all four algorithms have been implemented in MATLAB, C/C++ using Eigen and CUDA-C/C++.

### 4.1. MATLAB Implementations

All algorithms (PHD, MUSIC, EV & MN) have been implemented in MATLAB in double precision (default setting) and no special function under any toolbox has been used in the code flows.

### 4.2. C/C++ Implementations

In the C/C++ implementations, the *Eigen C++ template library* [16] is used since it provides a number of linear algebra utilities (various data-structure types, numerical solvers, etc.) with built-in optimized explicit vectorization (via different SIMD instruction sets activated by compiler options). Computation of Singular Value Decomposition (SVD) has a very significant role in determining the noise subspace for $\widehat{R}$ matrix and hence in the DOA estimation performance. In C/C++, SVD is realized by the *JacobiSVD method* of the *Eigen library* [17].

Additionally, in order to achieve a fair comparison with the CUDA implementations, all the thread-works in each parallelizable (SPMD-compatible) region are distributed into multiple logical cores by appropriate *OpenMP* pragmas. Robust multi-threaded implementation with higher efficiency is realized by using special constructs *(critical)* and special clauses *(dynamic-schedule)*.

### 4.3. CUDA Implementations

To identify the time-consuming parts and parts having parallelization potential, the C/C++ implementations have been profiled. During this process, the design cycle proposed by NVIDIA [22] has been adopted as the main guide throughout the CUDA code development and performance optimization process. This design cycle consists of four stages: *assessment*, *parallelization*, *optimization*, and *deployment* (APOD).

First, C/C++ codes have been analyzed to identify the segments dominating the execution time. Since all four DOA algorithms are quite similar, the sequential MUSIC algorithm implemented in C/C++ was chosen for time duration analysis. Time distribution for scan range of [0:359] x [1:90] is given in Figure 2.

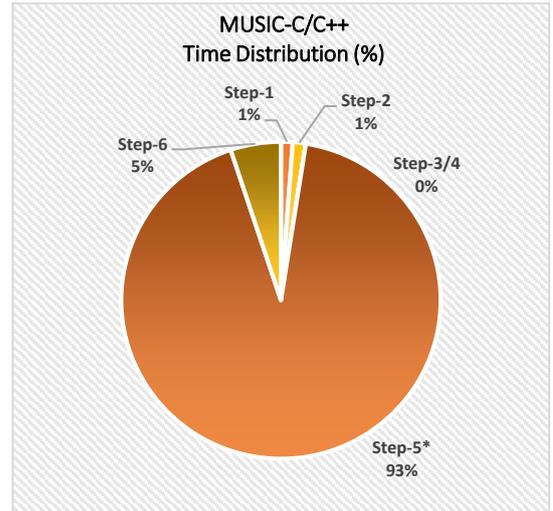

*Figure 2*: Time Distribution for C/C++ implementation of the MUSIC algorithm.

As seen in Figure 2, the hotspot in the overall execution is the pseudo-spectrum computation (Step-5*) taking up 93% of the whole runtime, which is Step-5 in Table 2 excluding steering vector computations. In order to verify the algorithm hotspot analytically, the total computation amount for each step is studied as in Table 4, where $M$, $N$, $D$ and $L$ are the number of sensors, samples, signals & scan angles, respectively. In our study, three variables other than $L$ are relatively small, whereas $L$ is swept in a broad range. As the amount of computation in Step-5 and 6 increases in a multiplicative relation with $L$, these parts are also analytically confirmed to be computationally dominant parts.

*Table 4:* Total computation amount at each algorithm step.

| Operation (Table 2) | Total Computation |
|---|---|
| Step-1 | $(3M^2N + MN)/2$ |
| Step-2 & Step-3 | $12M^3$ |
| Step-4 | $M^3 + (1/2 - D)M^2 - (1/2 + D)M$ |
| Step-5 | $L(2N^2 + N)$ |
| Step-6 | $L \log(L)$ |

As the next stage in the CUDA design cycle, the most time-dominant algorithm part (Step-5 in Table 2) is evaluated in terms of parallelization and the CPU-GPU hybrid structure is constructed as in Figure 3 for DOA CUDA implementations. As shown in Figure 3, GPU is utilized in Step-5.2 to Step-5.4 and optionally in Step-2. Parallelization of pseudo-spectrum computation is realized by kernelizing the corresponding steps, assigning each CUDA threads for computing spectrum value at a specific scan angle pair and enabling data flow between host (CPU) & device (GPU) side appropriately. Apart from this, in order to investigate the availability of an efficient GPU-based SVD computation, some numerical & performance experiments have been realized using cuSOLVER API (described in Section 5).

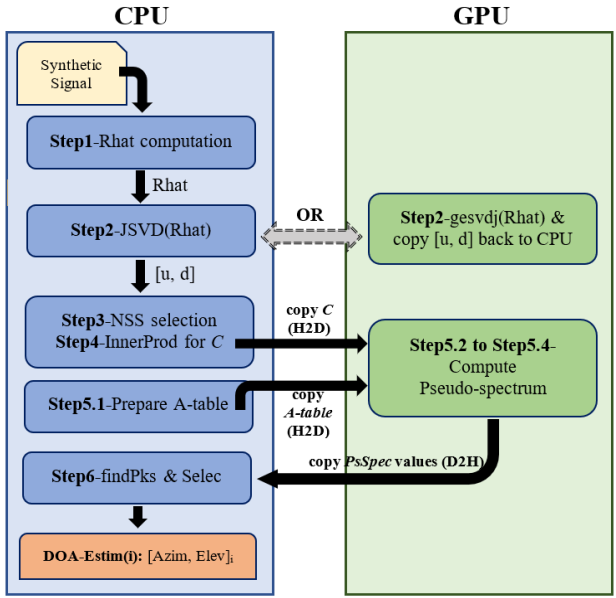

*Figure 3*: Heterogeneous CPU-GPU structure.

## 5. Numerical Validation and Performance Analysis

A specific DOA test scenario has been designed (Figure 4) for numerical validation and performance evaluation of the algorithms. The scenario assumes that two RF sources are uncorrelated, incoming signals are narrowband and carried at 15 MHz with 15 dB SNR level. Eight omnidirectional antennas are positioned in a uniformly circular way with a radius of 10 m.

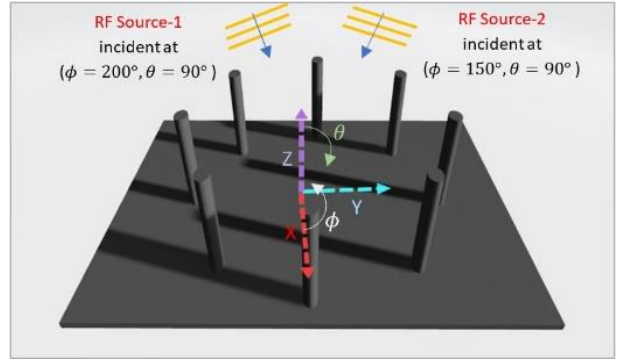

*Figure 4*: Test scenario setting including two sources.

### 5.1. General Numerical Accuracy Analysis

In order to validate the numerical accuracy of C/C++ and CUDA implementations of DOA methods in single precision (float32), resultant pseudospectrum values of the methods have been used as measurement data. The scan has been realized in azimuth = [0:1:359] and elevation = [90] (constant), hence 360 values are used for percent error measure against MATLAB (double precision) results. The error calculation is given in Eq. 4.

$$e_{percent} = \frac{100}{N} * \sum_{i=1}^{N} \frac{|y_{meas,i} - y_{gndTruth,i}|}{|y_{gndTruth,i}|} \quad (4)$$

*Table 5:* Numerical (percent) error comparison.

| DOA Methods | Ground Truth | Percent (%) Error | |
|---|---|---|---|
| | | C/C++ (float32) | CUDA (float32) |
| PHD | MATLAB (float64) | 0.016952 | 0.024270 |
| MUSIC | | 0.001229 | 0.001134 |
| EV | | 0.001720 | 0.002010 |
| MN | | 0.000850 | 0.000403 |

As seen in Table 5, the error in single precision (float32) implementations is at most 0.024270%. Since DOA estimation depends upon distinguishability of spectrum values from each other, the attained error is acceptable and verified by observing estimation results (all estimate DOA correctly).

### 5.2. Numerical and Performance Evaluation for SVD

In this experiment, in order to investigate the most suitable resource (CPU or GPU) and method for SVD computation, numerical and performance-based comparisons have been done between Eigen-JacobiSVD [17] (CPU) & cuSOLVER-gesvdj (GPU) [19]. SVD computation using cuSOLVER has been adapted from an official NVIDIA code example [8].

The residual errors are computed based on Eq. 5, where **A** is input matrix and **S**, **U** & **V** are singular values, Left & Right singular vectors, respectively.

$$residualErr = |A - U * S * V^H| \quad (5)$$

Table 6: Residual error comparison (for 8x8 matrix).

|  | **MATLAB** | **Eigen** | **cuSOLVER** |
|---|---|---|---|
| **Res.Error** | 1.075e-14 | 1.882e-05 | 6.391e-08 |

Table 7: SVD performance comparison (ms).

| Matrix Size: | MATLAB-SVD: | Eigen-JacobiSVD: | cuSOLVER-gesvdj: |
|---|---|---|---|
| 8x8 | 0.018 | 0.112 | 390.279 |
| 64x64 | 0.836 | 29.910 | 386.242 |
| 512x512 | 64.283 | 18651 | N/A |

Table 8: MATLAB and C/C++ single and multi-threaded performance comparison for different ranges (ms) for MUSIC.

| Scan Range | MATLAB | C/C++ single | C/C++ multi |
|---|---|---|---|
| **360 x 1** | 6.4 | 0.486 | 0.408 |
| **360 x 30** | 135.2 | 3.106 | 1.242 |
| **360 x 60** | 269.9 | 5.612 | 2.111 |
| **360 x 90** | 401.9 | 8.318 | 2.862 |

As can be seen from Table 6 and Table 7, while both methods are numerically suitable, the Eigen-based SVD method has a significantly better computational performance. Hence, the Eigen-JacobiSVD method has been selected for the SVD computation in CUDA codes.

### 5.3. MATLAB and C/C++ Performance Evaluation

As mentioned earlier, all DOA methods are similar in terms of code structure and time performance. Therefore, a single algorithm (MUSIC) has been used in initial performance comparisons.

Since an array manifold lookup table is created and reused in a continuously working virtual system, the time required for preparing this table is not included in the analysis and it can be assumed as a start-up overhead. Table 8 shows the results for MATLAB, single-threaded, and multi-threaded C/C++ implementations. The experiments show that C/C++ implementation provides a significant speedup against the MATLAB implementation with up to 140.43x at 360 x 90 scan range.

### 5.4. CUDA Performance Optimizations

After the initial experiments, further CUDA optimizations have been investigated following the best practices guideline [19] and the optimization steps in Table 9 have been applied consecutively. Steps 1 and 2 in Table 9 improves the kernel performance by 2.4x (reducing the execution time from 0.129 ms to 0.055 ms) and Step 3 improves the Device-to-Host (D2H) memcpy performance 1.35x (reducing the copy time from 0.098 ms to 0.072 ms).

### 5.5. CUDA Performance Evaluation on Different Platforms

After the optimizations, code performances have been measured on two system configurations. Config-1 with Intel i7-9750H 2.6 GHz (6 cores) CPU and NVIDIA GTX 1660TI GPU and Config-2 with Intel i7-7700HQ 2.8 GHz (4 cores) CPU and NVIDIA GTX 1050 GPU. The effect of differences in technical specs on the optimized CUDA code performance of the MUSIC algorithm is observed by measuring the total code duration for each scan range value, as shown in Table 10. As seen in this table, MUSIC-CUDA on Config-1 is about 1.5 times faster than one on Config-2, which is in line with expectations considering their technical specifications. The GPU version does not have an advantage at low scan ranges due to low utilization. However, in real-time DOA estimation applications, to cover all possible signal arrival directions, higher scan range values in the algorithms play an important role in the estimation accuracy. The experiments show that the speedup increases when the scan range is higher and up to 3.1x speedup can be achieved against the multi-threaded C/C++ implementation at 360 x 90 scan range.

Table 9: CUDA performance optimization steps.

| | |
|---|---|
| 1 | Since 8x8 array C is small in size (512 bytes) and its information is used by thousands of threads, instead of keeping this on global memory, it is kept on local memory with a capacity of 48 KB. |
| 2 | By declaring the pointer A_table with additional **const** and **restrict** qualifiers, some compiler-level optimization is enabled. If the CUDA compiler decides that data on the memory pointed by a specific pointer is read-only throughout the kernel lifetime, it enables access to this data via a read-only cache mechanism. |
| 3 | By allocating some host arrays (C, P-table) in the pinned (page-locked) memory, corresponding data is not paged out of the physical memory. This makes data transfers between host and device more efficient. |

Table 10: GPU Performance comparison on different system configurations (ms) and speedup against multi-threaded C/C++ implementation for MUSIC.

| Scan Range | GPU Config-1 | | GPU Config-2 | |
|---|---|---|---|---|
| | Runtime | Speedup | Runtime | Speedup |
| **360 x 1** | 0.438 | 0.93x | 0.677 | 0.60x |
| **360 x 30** | 0.647 | 1.92x | 1.017 | 1.22x |
| **360 x 60** | 0.812 | 2.60x | 1.257 | 1.68x |
| **360 x 90** | 0.924 | 3.10x | 1.447 | 1.98x |

## 6. Conclusions

In this study, four noise subspace-based super-resolution DOA algorithms (PHD, MUSIC, EV and MN) have been studied with regards to their numerical accuracy and performance on CPU and GPU. The parallelization of the algorithms on GPU has been realized with particular attention to maintain their numerical accuracy. The experiments showed that the optimized CUDA code can provide a speedup of 3.1x compared to the multi-threaded CPU code.

As a future work, the last step in the code flow (Step-6 in Figure 3) could be realized by *max-reduction* operation utilizing shared memory to obtain further performance improvements compared to existing Eigen library-based operations.

## 7. References

[1] T. E. Tuncer and B. Friedlander, Classical and modern direction-of-arrival estimation, Academic Press, 2009.

[2] H. Krim and M. Viberg, "Two decades of array signal processing research: the parametric approach," *IEEE Signal Processing Magazine,* vol. 13, no. 4, pp. 67-94, 1996.

[3] F. Yan, M. Jin and X. Qiao, "Low-Complexity DOA Estimation Based on Compressed MUSIC and Its Performance Analysis," *IEEE Transactions on Signal Processing,* pp. 1915 - 1930, 15 April 2013.

[4] Z. Chen, G. K. Gokeda and Y. Yu, Introduction to Direction-of-Arrival Estimation, Artech House, 2010.

[5] V. F. Pisarenko, "The retrieval of harmonics from a covariance function," *Geophysical Journal International,* vol. 33, no. 3, pp. 347-366, 1973.

[6] M. H. Hayes, Statistical digital signal processing and modeling, John Wiley & Sons, 2009.

[7] R. Schmidt, "Multiple emitter location and signal parameter estimation," *IEEE transactions on antennas and propagation,* vol. 34, no. 3, pp. 276-280, 1986.

[8] D. Johnson and S. DeGraaf, "Improving the resolution of bearing in passive sonar arrays by eigenvalue analysis," *IEEE Transactions on Acoustics, Speech, and Signal Processing,* vol. 30, no. 4, pp. 638 - 647, 1982.

[9] R. Kumaresan and D. W. Tufts, "Estimating the Angles of Arrival of Multiple Plane Waves," *IEEE Transactions on Aerospace and Electronic Systems,* Vols. AES-19, no. 1, pp. 134-139, 1983.

[10] OMP Architecture Review Board, "The OpenMP API," 8 November 2018. [Online]. Available: https://www.openmp.org/.

[11] Z. Zou, W. Hongyuan and Y. Guowen, "An Improved MUSIC Algorithm Implemented with High-speed Parallel Optimization for FPGA," in *7th International Symposium on Antennas, Propagation & EM Theory*, Guilin, China, 2006.

[12] M. W. Majid, T. E. Schmuland and M. M. Jamali, "Parallel implementation of the wideband DOA algorithm on single core, multicore, GPU and IBM cell BE processor," *Science Journal of Circuits, Systems and Signal Processing,* vol. 2, no. 2, pp. 29-36, 2013.

[13] M. W. Maajid, G. Mirzaei and M. M. Jamali, "Parallel Wavelet Transform Based Wideband Direction-of-Arrival (DOA) on Multicore/GPU," in *IEEE International Conference on Electro-Information Technology, EIT*, Rapid City, SD, USA, 2013.

[14] Z. Lu, L. Zhang, J. Zhang and J. Zhang, "Parallel optimization of broadband underwater acoustic signal MUSIC algorithm on GPU platform," in *2017 4th International Conference on Systems and Informatics (ICSAI)*, Hangzhou, China, 11-13 Nov. 2017.

[15] B. Ottersten, M. Viberg, P. Stoica and A. Nehorai, "Exact and Large Sample Maximum Likelihood Techniques for Parameter Estimation and Detection in Array Processing," Springer, Berlin, Heidelberg, 1993, pp. 99-151.

[16] Inria-TUX Family, "Eigen C++ template library," Inria, 11 December 2018. [Online]. Available: http://eigen.tuxfamily.org/index.php

[17] Inria-TUX Family, "Eigen::JacobiSVD," 2014. [Online]. Available: https://eigen.tuxfamily.org/dox/classEigen_1_1JacobiSVD.html.

[18] NVIDIA Corporation, "CUDA C++ Best Practices Guide," [Online]. Available: https://docs.nvidia.com/cuda/cuda-c-best-practices-guide/index.html.

[19] NVIDIA Corporation, "gesvdj-example-1 (G.2)," 28 November 2019. [Online]. Available: https://docs.nvidia.com/cuda/cusolver/index.html#gesvdj-example1.

[20] MathWorks, "MATLAB-SVD," MathWorks, 2006. [Online]. Available: https://www.mathworks.com/help/matlab/ref/double.svd.html.

[21] S. Cook, CUDA Programming: A Developer's Guide to Parallel Computing with GPUs, Newnes, 2012.